\documentclass[11pt]{article}
\pdfoutput = 1
\usepackage[utf8]{inputenc}
\usepackage{color,graphicx}
\usepackage{epsfig}
\usepackage{amsmath}
\usepackage{verbatim}
\usepackage{amssymb}
\usepackage{mathabx}
\usepackage{stmaryrd}
\usepackage{empheq}
\usepackage{simpler-wick}
\usepackage{esint}
\usepackage{physics} 
\usepackage{amsfonts}
\usepackage{array}
\usepackage{setspace}
\usepackage{xcolor}
\usepackage{braket}
\usepackage{float}
\usepackage{url}
\usepackage{mathtools}
\usepackage{tensor}
\usepackage{bbold}
\usepackage{slashed}
\usepackage{tikz}
\usetikzlibrary{decorations.markings}
\usetikzlibrary{decorations.pathmorphing}
\usepackage{tikz,pgfplots}
\usetikzlibrary{tikzmark,calc}
\usepackage{epstopdf}
\usepackage{subfigure}
\usepackage[labelfont=bf]{caption}
\usepackage{pgfplots}
\usepackage{mathrsfs}
\usepackage{fancybox}
\usepackage{eurosym}
\usepackage{tcolorbox}
\usepackage{tensor}
\allowdisplaybreaks
\usepackage{changepage}
\usepackage[mathcal]{eucal}

\usepackage[margin = 2.2cm]{geometry}
\usepackage[ragged]{footmisc}
\usepackage[maxbibnames=99,
    sorting=none,
    backend=biber,
    style=numeric,
]{biblatex}
\usepackage[bookmarks=true,bookmarksnumbered=false,hyperindex=true,bookmarksopen=true,hyperfigures=true,colorlinks=true,linkcolor=webblue,citecolor=webgreen,urlcolor=webblue,breaklinks]{hyperref}
\definecolor{webgreen}{rgb}{0, 0.5, 0}
\definecolor{webblue}{rgb}{0, 0, 0.5}
\definecolor{webred}{rgb}{0.5, 0, 0}
\definecolor{darkgreen}{rgb}{0,0.5,0}

\def\ben{\begin{equation}}
\def\een{\end{equation}}

\numberwithin{equation}{section}

\def\spc{\hspace{1pt}}

\newcommand{\marknode}[2]{\tikzmarknode{#1}{#2}}

\newcommand{\contractaa}[5][]{%
\begin{tikzpicture}[overlay, remember picture]
\coordinate (L) at ($(pic cs:#2)+(0,2ex)$);
\coordinate (R) at ($(pic cs:#3)+(0,2ex)$);
\coordinate (Lt) at ($(L)+(0,#4)$);
\coordinate (Rt) at ($(R)+(0,#4)$);
\coordinate (#5) at ($(Lt)!0.5!(Rt)$);
\draw[#1]
  (L) -- (Lt) -- (Rt) -- (R);
\end{tikzpicture}%
}

\begin{document}

\thispagestyle{empty}
\begin{adjustwidth}{-1cm}{-1cm}
\begin{center}
    ~\vspace{9mm}
    
     {\LARGE \bf 
   Cosmology of the Baby Universe in Coupled SYK
   }
   \end{center}
    \end{adjustwidth}
    \begin{center}
   \vspace{0.4in}
    
    {\bf Andrew Sontag}

    \vspace{0.4in}
    {
    Department of Physics, Princeton University, Princeton, NJ 08544, USA
    }
    \vspace{0.1in}
    
    {\tt sontaga@princeton.edu}
\end{center}

\vspace{0.4in}

\begin{abstract}

We analyze the structure of the baby universe which arises in the coupled SYK realization of the AS$^2$ geometry. To do this, we develop and leverage combinatorial arguments about traces of many Majoranas which remain valid at finite $N$ and $p$ and thus furnish the finite $\lambda$ completion of the chord rules found in \cite{Sontag:2026iiu}. We find that there is a first order phase transition which persists in the $\beta\rightarrow\infty$ limit between the thermal AdS and AS$^2$ phases parameterized by the mass of the heavy particle insertion and the ratio between the SYK and Maldacena-Qi coupling constants. We then explore Lorentzian time evolution in the baby universe and find a cosine scale factor confirming the big bang / big crunch nature of the bulk cosmology.

\end{abstract}

\def\nspc{\hspace{-1pt}}
\pagebreak
\setcounter{page}{1}
\tableofcontents

\section{Introduction}\label{sect:intro}

Understanding how semiclassical closed universe cosmologies can arise from holographic formulations of quantum gravity has proven to be a difficult first step towards a unified theory of our own universe's beginning and fate. While de Sitter quantum gravity remains in its infancy \cite{Narovlansky:2025tpb,Collier:2025lux,Tietto:2025oxn,marini20263dneardesittergravity,blommaert2026observersquantization3dsitter,banks2026doesmeanquantumgravitational,fujiki2026sitterholographylorentziantorus,cotler2026innerproductscloseduniverses}, many have chosen instead to study closed patches of Anti-de Sitter space \cite{Antonini:2023hdh,Antonini:2024mci,Sasieta:2025vck,Antonini:2025ioh} using holographic dictionaries that are well-established for open spacetimes \cite{Maldacena:1997,Witten:1998qj,Gubser:1998bc}. This has spurred many questions as to whether these kinds of prescriptions can really generate well-behaved closed universes \cite{Gesteau:2025obm,Engelhardt:2025vsp,Kudler-Flam:2025cki,liu2025filteringcftslargen} and if so, what kinds of Hilbert spaces \cite{McNamara:2020uza,Usatyuk_2024,usatyuk2025closeduniversesfactorizationensemble,Mori:2025jej} and quantum properties \cite{Marolf:2020xie,Harlow:2025pvj,Liu:2025cml} those universes might have.

Perhaps the most elucidating example of this line of research is the study of the AS$^2$ geometry \cite{Antonini:2023hdh}, which contains a closed section called the baby universe, using the duality between quantum gravity in AdS$_2$ and the double-scaled SYK (DSSYK) model \cite{Maldacena:2016hyu,Kitaev:2017awl,Berkooz:2018qkz,Berkooz:2018jqr,Goel:2023svz} with a Maldacena-Qi (MQ) coupling \cite{Maldacena:2018lmt}. It was recently shown \cite{Sontag:2026iiu} that DSSYK-MQ indeed has saddle points which correspond to the AS$^2$ bulk geometry and other related topologies, and that these saddle points admit well-behaved chord diagram descriptions and thus semiclassical bulk Hilbert spaces \cite{Lin_2022,Lin:2023trc}. By studying the properties of these saddle points and their closed sections, we can gain a better understanding of what ingredients are required for a well-behaved quantum theory of cosmology, and of what goes wrong should we omit them.

In this paper, we study the DSSYK-MQ AS$^2$ baby universe by looking directly at traces of large numbers of Majorana fermions.
This expands upon and solidifies previous $G\Sigma$ results in DSSYK-MQ \cite{Sontag:2026iiu} by honing in on nonperturbative finite-$N$ effects.
We start with a review of the SYK model and the MQ coupling in section \ref{subsect:introSYKMQ}, followed by a review of the $G\Sigma$ approach and the AS$^2$ geometry in section \ref{subsect:introGS}. We then introduce our Majorana trace methods in section \ref{sect:wohp} to compute finite $\lambda$ two-point and four-point correlators in the unperturbed DSSYK-MQ model, and we show that in the correct limit these reduce to the answers found in \cite{Sontag:2026iiu}. In section \ref{sect:whp} we insert the heavy particle to generate an AS$^2$ geometry and study the proliferation of wormhole chords, finding a first order phase transition between a perturbed thermal AdS and an AS$^2$ saddle. In section \ref{sect:big_bang_crunch} we look at Lorentzian time evolution inside and outside the baby universe, and we consider a prescription for the former which generates a cosine scale factor matching the expected big bang / big crunch cosmology. Appendix \ref{app:FourierWalk} contains the main mathematical workhorse behind our results, a Fourier transform on $(\mathbb{Z}_2)^N$. Appendix \ref{app:B} details a perpendicular numerical method for computing the same observables that we use to check the validity of our double scaling approximations.

While we were finalizing this manuscript, another work \cite{bettaque2026sykwormholemicroscope} was published applying the same mathematical methods in Appendix~\ref{app:FourierWalk} to a different but related setup. We provide a comparison of our results in the discussion.

\subsection{Review of SYK and MQ}\label{subsect:introSYKMQ}

The SYK model consists of $N$ Majorana fermions $\{\psi_i,\psi_j\}=\delta_{ij}$ evolving under a $p$-local Hamiltonian with Gaussian random couplings
\begin{equation}\label{eq:intro_SYK_Hamiltonian}
H_{\text{SYK}}=i^{p/2}\sum_{1\leq i_1<...<i_p\leq N}J_{i_1...i_p}\psi_{i_1}...\psi_{i_p}
\end{equation}
\begin{equation}\label{eq:intro_SYK_J_distribution}
\langle J_{i_1...i_p}^2\rangle\, =\, \frac{2^{p-1} N }{p^2 \spc \Bigl(  {{N}\atop p}  \Bigr)}\,{\mathcal{J}}^2
\end{equation}
where $\mathcal{J}$ is a constant with units of energy. The double scaling limit takes $N$ and $p$ both to infinity while holding $\lambda=2p^2/N$ fixed. We are interested in computing thermal matter correlators
\begin{equation}\label{eq:intro_SYK_thermal_correlator}
    \big\langle\mathbb{O}(\tau)\mathbb{O}(0)\big\rangle=\frac{1}{Z_{\text{SYK}}(\beta)}\Tr\Big[e^{-(\beta-\tau)H_{\text{SYK}}}\,\mathbb{O}^\dagger\;e^{-\tau H_{\text{SYK}}}\,\mathbb{O}\Big]
\end{equation}
with
\begin{equation}\label{eq:intro_SYK_matter}
    \mathbb{O}=i^{p'/2}\sum_{1\leq i_1<...<i_{p'}\leq N}K_{i_1...i_{p'}}\psi_{i_1}...\psi_{i_{p'}}\,,\quad\;\;\frac{p'}{p}=\Delta
\end{equation}
where the $K_{i_1...i_{p'}}$ are distributed analogously to but independently of the $J_{i_1...i_p}$ as
\begin{equation}
    \langle K_{i_1...i_{p'}}^2\rangle\, =\, \frac{2^{p'-1} N }{p^2 \spc \Bigl(  {{N}\atop p'}  \Bigr)}\,
  {\mathcal{K}}^2.
\end{equation}

Taking the low energy limit $\beta\mathcal{J}\rightarrow\infty$ and integrating out the randomness in $J_{i_1...i_p}$ and $K_{i_1...i_{p'}}$ allows these correlators to probe geodesic distances $\ell(\tau)$ between points separated by Euclidean distance $\tau$ on the boundary of a static AdS$_2$ disk of the same inverse temperature $\beta$ \cite{harlow2019factorizationproblemjackiwteitelboimgravity}. The holographic dictionary is
\begin{equation}\label{eq:intro_holographic_dictionary_SYK}
\big\langle\mathbb{O}(\tau)\mathbb{O}(0)\big\rangle\,\longleftrightarrow\,\Big\langle e^{-\Delta\ell(\tau)}\Big\rangle
\end{equation}
where the left side is the boundary correlator from \eqref{eq:intro_SYK_thermal_correlator} and the right side is a path-integrated bulk observable in $D=2$ quantum gravity. This is additionally useful because the DSSYK correlator permits a chord diagram expansion which can be interpreted as furnishing a semiclassical bulk Hilbert space, see \cite{Berkooz:2024lgq} for an introduction.

Motivated by recent progress in resolving bulk topology transitions \cite{Sontag:2026iiu,Sasieta:2025vck,aguilargutierrez2026qaskeydeformationsdoublescaledsyk}, we consider a pair of SYK models coupled by a Maldacena-Qi interaction \cite{Maldacena:2018lmt}. This DSSYK-MQ model consists of two sets of $N$ Majorana fermions $\{\psi_i^{(0,1)}\}$ with Hamiltonian
\begin{equation}\label{eq:intro_H_MQ}
    H_{MQ}=H_{SYK}^{(0)}+H_{SYK}^{(1)}+i\frac{\mu}{p}\sum_{i=1}^N \psi_i^{(0)}\psi_i^{(1)}
\end{equation}
where
$H_{SYK}^{(a)}$ is a copy of \eqref{eq:intro_SYK_Hamiltonian} labeled by $a\in\{0,1\}$. We will set the couplings on each site equal $J_{i_1...i_p}^{(1)}= J_{i_1...i_p}^{(0)}$ and take $p\equiv 0 \text{ mod }4$. The analogous holographic dictionary now involves two circular boundaries, and bulk geodesics may start and end on either one. We denote those that start on boundary $a$ and end on boundary $b$ as ``$ab$-type.'' We define length variables $\ell_{ab}(\tau)$ and write
\begin{equation}\label{eq:intro_holographic_dictionary_MQ}
\big\langle\mathbb{O}^{(a)}(\tau)\mathbb{O}^{(b)}(0)\big\rangle\,\longleftrightarrow\,\Big\langle e^{-\Delta\ell_{ab}(\tau)}\Big\rangle
\end{equation}
where $a$ and $b$ are each in $\{0,1\}$.\footnote{In situations where the geodesic length is not measured equally by particles with very different masses $\Delta$, we will use the variables $\ell$ and $\ell_{ab}$ to mean the length measured by particles with $\Delta=\mathcal{O}(1)$.}

We will find it helpful moving forward to choose a matrix representation for these Majoranas. One convenient choice is to pair each $\psi_i^{(0)}$ with $\psi_i^{(1)}$ to create a Dirac fermion, which we can represent with a string of $N$ Pauli matrices. We thus define:
\begin{align}
    \psi_i^{(0)}=\frac{1}{\sqrt{2}}\Bigg(\bigotimes_{j=1}^{i-1}\sigma_z\Bigg)\otimes\sigma_x\,\otimes\Bigg(\bigotimes_{j=i+1}^{N}\mathbb{1}\Bigg)\label{eq:intro_JorWig_x}\\
    \psi_i^{(1)}=\frac{1}{\sqrt{2}}\Bigg(\bigotimes_{j=1}^{i-1}\sigma_z\Bigg)\otimes\sigma_y\,\otimes\Bigg(\bigotimes_{j=i+1}^{N}\mathbb{1}\Bigg)\label{eq:intro_JorWig_y}
\end{align}
Observe that this indeed satisfies $\{\psi_i^{(a)},\psi_j^{(b)}\}=\delta_{ij}\,\delta_{ab}$ as required.

\subsection{Review of $G\Sigma$ and AS$^2$}\label{subsect:introGS}

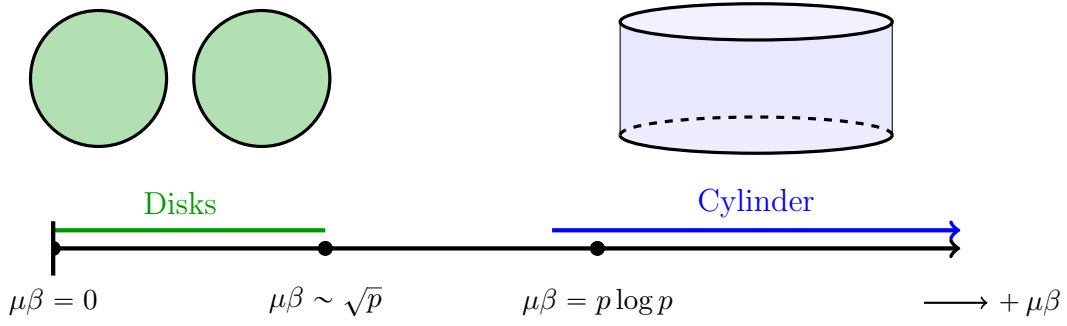
\begin{figure}[t]
\centering
\begin{tikzpicture}[xscale=1.2,yscale=-1.2]

\draw[ultra thick, ->] (0,0) -- (10,0);
\draw[thick, ->] (9.6,.6) -- (10.3,.6) node[right]{$+\,\mu\beta$};

\foreach \x/\lab in {0/$\mu\beta=0$, 3/$\mu\beta\sim\sqrt{p}$, 6/$\mu\beta=p\log p$} {
    \node[fill=black, circle, inner sep=2pt] at (\x,0) {};
    \node[yshift=-20] at (\x,0) {\lab};
}
\draw[ultra thick, green!60!black] (0,-0.2) -- (3,-0.2);
\node[yshift=17, green!60!black] at (1.4,0) {\large Disks};
\draw[ultra thick, blue, ->] (5.5,-0.2) -- (10,-0.2);
\node[yshift=17, blue] at (7.75,0) {\large Cylinder};

\fill[white]
(0,-0.3) -- (-0.3, -0.3) -- (-0.3, 0.3) -- (0, 0.3) -- cycle;
\draw[ultra thick] (0,-0.3) -- (0,0.3);

\draw[very thick, fill=green!60!black!30] (0.5,-1.875) circle (0.75);
\draw[very thick, fill=green!60!black!30] (2.3,-1.875) circle (0.75);

\fill[blue!9]
(7.75-1.5,-1.25)
arc[start angle=180,end angle=360,x radius=1.5,y radius=0.2]
-- (7.75+1.5,-2.5)
arc[start angle=360,end angle=180,x radius=1.5,y radius=0.2]
-- cycle;
\draw[very thick, fill=blue!5] (7.75,-2.5) ellipse [x radius=1.5, y radius=0.2];

\draw[very thick, fill=blue!7] (7.75-1.5,-1.25) arc[start angle=180,end angle=0,x radius=1.5,y radius=0.2];
\draw[very thick, dashed, fill=blue!7] (7.75+1.5,-1.25) arc[start angle=360,end angle=180,x radius=1.5,y radius=0.2];

\draw (7.75-1.5,-1.25) -- (7.75-1.5,-2.5);
\draw (7.75+1.5,-1.25) -- (7.75+1.5,-2.5);

\end{tikzpicture}
\caption{Diagram representing which regions of inverse temperature $\beta$ correspond to which geometric bulk phases of the unperturbed DSSYK-MQ model. Infinite temperature ($\beta=0$) is on the left edge. Each phase has two circular boundaries corresponding to the two SYK sites. For further discussion of these phases see \cite{Maldacena:2018lmt} and \cite{Sontag:2026iiu}.}
\label{fig:intro_beta_number_line}
\end{figure}

The DSSYK-MQ model admits a helpful mean-field theory approach parameterized by the bilocal fields
\begin{equation}\label{eq:intro_define_Gab}
    G_{ab}(\tau,\tau')=\frac{(-i)^{(a+b)^2}}{N}\sum_{i=1}^N \psi_i^{(a)}(\tau)\psi_i^{(b)}(\tau')
\end{equation}
and their Lagrange multipliers $\Sigma_{ab}(\tau,\tau')$. In the double scaling limit, we can substitute $\langle(2G_{ab}(\tau,\tau'))^p\rangle$ for $\langle\mathbb{O}^{(a)}(\tau)\mathbb{O}^{(b)}(\tau')\rangle$ in \eqref{eq:intro_holographic_dictionary_MQ} to probe the bulk geometry. Using this method, it was found in \cite{Maldacena:2018lmt} that the bulk dual of this model undergoes a Hawking-Page transition at $\mu\beta=\mathcal{O}(p\ln p)$ from a pair of Euclidean disks at high temperature corresponding to a black hole spacetime (green in Figure~\ref{fig:intro_beta_number_line}) to a cylinder at low temperature corresponding to thermal AdS without a black hole (blue in Figure~\ref{fig:intro_beta_number_line}). Additionally, the actions of these saddle points include a term which looks like $S_0\,\chi$ where $S_0=\frac{N}{2}\ln 2$ is the maximum entropy of a single SYK site and $\chi$ is the Euler characteristic of the bulk Euclidean space ($2$ for the disks and $0$ for the cylinder).

It was then found in \cite{Sontag:2026iiu} using the same mean-field approach that these two saddle points each admit a well-behaved chord diagram description in the small $\lambda$ regime and thus a semiclassical bulk Hilbert space. These chord Hilbert spaces are semiclassical in the sense that they arise in the $N\rightarrow \infty$ limit and cannot see finite $N$ effects like bulk topology change. The full UV complete Hilbert space is that of the boundary DSSYK-MQ model, which reduces to an apparent bulk chord space far from topology-changing conditions.

\begin{figure}[h!]
\vspace{10pt}
\centering
\begin{tikzpicture}[scale=.54]
\def\rout{3.6}
\def\rin{0.5}
\def\br{1.8} 
\def\g{2.8}
\def\h{4.2}
\def\spac{15}
\fill[violet!12]
(-\br-\rin-\rout,0)
arc[start angle=180,end angle=-180,x radius=\rout,y radius=\rout] -- cycle;
\fill[violet!12]
(\br+\rin+\rout,0)
arc[start angle=360,end angle=0,x radius=\rout,y radius=\rout] -- cycle;

\draw[thick, dashed] (-\br-\rin-\rout,0) -- (\br+\rin+\rout,0);

\fill[white]
(-\br-2*\rin,0)
arc[start angle=-180,end angle=180,x radius=\rin,y radius=\rin] -- cycle;
\fill[white]
(\br+2*\rin,0)
arc[start angle=0,end angle=360,x radius=\rin,y radius=\rin] -- cycle;

\draw[very thick] (-\br-\rin-\rout,0) arc[start angle=180,end angle={360-acos((\rin+\br)/\rout)},x radius=\rout,y radius=\rout];
\draw[very thick] (\br+\rin+\rout,0) arc[start angle=360,end angle={180+acos((\rin+\br)/\rout)},x radius=\rout,y radius=\rout];
\draw[very thick] (-\br-\rin-\rout,0) arc[start angle=180,end angle={0+acos((\rin+\br)/\rout)},x radius=\rout,y radius=\rout];
\draw[very thick] (\br+\rin+\rout,0) arc[start angle=0,end angle={180-acos((\rin+\br)/\rout)},x radius=\rout,y radius=\rout];
\draw[thick] (-\br-2*\rin,0) arc[start angle=-180,end angle=180,x radius=\rin,y radius=\rin];
\draw[thick] (\br+2*\rin,0) arc[start angle=360,end angle=0,x radius=\rin,y radius=\rin];

\draw[brown, line width=3pt]
    ({0}, {-1*sqrt(\rout^2 - (\br+\rin)^2)}) node[fill=brown, circle, inner sep=3pt] {}
    -- (0,\g) node[fill=brown, circle, inner sep=3pt] {};
\node at ({0.5}, {-1*sqrt(\rout^2 - (\br+\rin)^2)+0.2}) {\ \ \ \Large $\mathbb{O}$};
\node at ({0.5}, -{-1*sqrt(\rout^2 - (\br+\rin)^2)-0.2}) {\ \ \ \Large $\mathbb{O}$};
\fill[yellow!10]
(-\br-\rin-\rout+\spac,-\h) -- (-\br-\rin-\rout+\spac,\h) -- (-\br-2*\rin+\spac,\h) -- (-\br-2*\rin+\spac,-\h) -- cycle;
\fill[yellow!10]
(\br+\rin+\rout+\spac,-\h) -- (\br+\rin+\rout+\spac,\h) -- (\br+2*\rin+\spac,\h) -- (\br+2*\rin+\spac,-\h) -- cycle;
\fill[yellow!10]
(-\br+\spac,0) to[out=90, in=225] (-0.5+\spac,0.9*\h) decorate[decoration={zigzag, amplitude=2pt, segment length=4.2pt}]{ -- (0.5+\spac,0.9*\h)} to[out=315, in=90] (\br+\spac,0) -- cycle;
\fill[yellow!10]
(-\br+\spac,0) to[out=270, in=135] (-0.5+\spac,-0.9*\h) decorate[decoration={zigzag, amplitude=2pt, segment length=4.2pt}]{ -- (0.5+\spac,-0.9*\h)} to[out=45, in=270] (\br+\spac,0) -- cycle;
\draw[brown, line width=3pt] (0+\spac,-0.95*\h) -- (0+\spac,0.95*\h);
\fill[white]
decorate[decoration={zigzag, amplitude=2pt, segment length=4.2pt}]{(-0.5+\spac,0.9*\h) -- (0.5+\spac,0.9*\h)} -- (0+\spac,\h) -- cycle;
\fill[white]
decorate[decoration={zigzag, amplitude=2pt, segment length=4.2pt}]{(-0.5+\spac,-0.9*\h) -- (0.5+\spac,-0.9*\h)} -- (0+\spac,-\h) -- cycle;

\draw[thick, dashed] (-\br-\rin-\rout+\spac,0) -- (\br+\rin+\rout+\spac,0);

\fill[white]
(-\br-2*\rin+\spac,0)
arc[start angle=-180,end angle=180,x radius=\rin,y radius=\rin] -- cycle;
\fill[white]
(\br+2*\rin+\spac,0)
arc[start angle=0,end angle=360,x radius=\rin,y radius=\rin] -- cycle;

\draw[very thick] (-\br-\rin-\rout+\spac,-\h) -- (-\br-\rin-\rout+\spac,\h);
\draw[very thick] (\br+\rin+\rout+\spac,-\h) -- (\br+\rin+\rout+\spac,\h);
\draw[thick] (-\br-2*\rin+\spac,-\h) -- (-\br-2*\rin+\spac,\h);
\draw[thick] (\br+2*\rin+\spac,-\h) -- (\br+2*\rin+\spac,\h);
\draw[thick] (-\br+\spac,0) to[out=90, in=225] (-0.5+\spac,0.9*\h) decorate[decoration={zigzag, amplitude=2pt, segment length=4.2pt}]{ -- (0.5+\spac,0.9*\h)} to[out=315, in=90] (\br+\spac,0);
\draw[thick] (-\br+\spac,0) to[out=270, in=135] (-0.5+\spac,-0.9*\h) decorate[decoration={zigzag, amplitude=2pt, segment length=4.2pt}]{ -- (0.5+\spac,-0.9*\h)} to[out=45, in=270] (\br+\spac,0);

\node at (-\br/3+\spac, 0.55) {\Large $\mathbb{O}$};

\medskip
\end{tikzpicture}
\caption{Diagrams representing the AS$^2$ geometry in Euclidean (left panel) and Lorentzian (right panel) signature. The thick black lines are asymptotic boundaries, and the thin black lines are places where the areal radius vanishes. The dashed black lines show the Cauchy slice shared between the two panels. The brown lines represent the worldline of the heavy particle $\mathbb{O}$. The Lorentzian patch containing $\mathbb{O}$ is the big bang / big crunch baby universe.
\label{fig:intro_ASS}}
\end{figure}
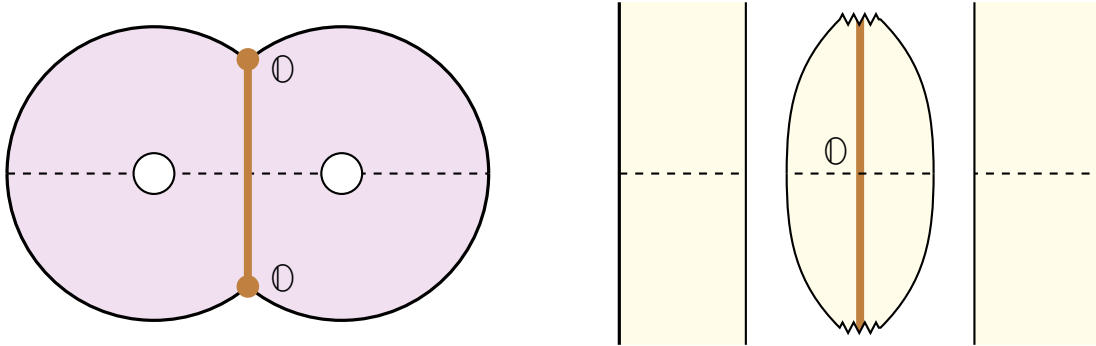

It was also found that the $G\Sigma$ action has yet another topologically-inequivalent saddle point corresponding to the AS$^2$ geometry \cite{Antonini:2023hdh} in which a pair of thermal AdS spaces are entangled with a closed big bang / big crunch cosmology known as a baby universe. This geometry is shown in Figure~\ref{fig:intro_ASS}. This becomes the dominant saddle point when a sufficiently heavy particle (brown line in Figure~\ref{fig:intro_ASS}) is inserted and annihilated on opposite sides of a thermal circle which would otherwise have been (well) below the Hawking-Page temperature. If the ``energetic'' cost of this particle having to travel all the way around the Euclidean AdS geometry outweighs the ``entropic'' cost of topology change, then a wormhole is created as a shortcut. In two dimensions, this gives us a cylinder with a handle, which has $\chi=-2$. Choosing a Cauchy slice (dashed line in Figure~\ref{fig:intro_ASS}) that cuts through this wormhole gives us a closed section, which if continued into Lorentzian signature generates a geometrically disconnected baby universe containing the heavy particle. On the DSSYK-MQ side, this heavy operator insertion changes the Hartle-Hawking state into a partially entangled thermal state (PETS) \cite{Goel:2018ubv}.

These findings left a number of open questions which the $G\Sigma$ approach was not well-equipped to easily answer: What are the chord rules at finite $\lambda$ and in the AS$^2$ geometry? How does tuning $\mathcal{J}$ affect the baby universe? Where can we see the $\chi=-2$ contribution to the action for the AS$^2$ saddle? To answer these and others, we need to make combinatorial arguments directly about traces of Majoranas.

\section{Without the heavy particle}\label{sect:wohp}

We start by analyzing the DSSYK-MQ partition function
\begin{equation}\label{eq:wohp_Z_MQ}
    Z_{MQ}[\beta]=\Tr\Big(e^{-\beta H_{MQ}}\Big)
\end{equation}
without a heavy particle insertion. This calculation serves both as a warm-up for the AS$^2$ case and to determine the finite $\lambda$ chord rules in the bulk. The first step is to treat $H_{MQ}$ in \eqref{eq:intro_H_MQ} as an interaction picture Hamiltonian whose ``free'' part is the Maldacena-Qi term $\mu H_\text{free}=i\frac{\mu}{p}\sum\psi^{(0)}\psi^{(1)}$ and whose ``interaction'' is $H^{(0)}_{SYK}+H^{(1)}_{SYK}$. We can then expand \eqref{eq:wohp_Z_MQ} as a Dyson series
\begin{equation}\label{eq:wohp_Dyson_series}
    Z_{MQ}[\beta]=\sum_{n=0}^\infty \frac{(-1)^n}{n!}\sum_{a_i\in\{0,1\}}\int_0^\beta d\tau_1...d\tau_n \Tr\Bigg(\mathcal{T}\bigg(e^{-\beta\,\mu H_\text{free}}\,H_{a_1}(\tau_1)...H_{a_n}(\tau_n)\bigg)\Bigg)
\end{equation}
where the $\mathcal{T}$ implements a Euclidean time-ordering from $\tau=0$ to $\beta$ and the $H_{a_i}$ is shorthand for $H_{SYK}^{(a_i)}$. Integrating out the Gaussian randomness in the SYK couplings $J_{i_1...i_p}$ lets us write this time-ordered quantity as a sum over pairwise contractions of the $H_a$'s. This forces $n$ to be even, so we can drop the $(-1)^n$. We can then consider just these contracted traces.

First consider the case with only one contracted pair. This takes the form
\begin{align}
&\Tr\bigg(\marknode{A}{H}_a\,e^{-\tau\,\mu H_\text{free}}\,\marknode{B}{H}_b\,e^{-(\beta-\tau)\,\mu H_\text{free}}\bigg)\contractaa[black, thick]{A}{B}{1.7ex}{}\label{eq:wohp_trace_traverse_ab}\\=\sum_{i_1...i_p}J_{i_1...i_p}^2&\Tr\bigg(\psi_{i_1}^{(a)}...\psi_{i_p}^{(a)}\,e^{-\tau\,\mu H_\text{free}}\,\psi_{i_1}^{(b)}...\psi_{i_p}^{(b)}\,e^{-(\beta-\tau)\,\mu H_\text{free}}\bigg).\label{eq:wohp_trace_traverse_ab_psis}
\end{align}
We can then replace $J_{i_1...i_p}^2$ with $\langle J_{i_1...i_p}^2\rangle$ and use our matrix representation \eqref{eq:intro_JorWig_x}-\eqref{eq:intro_JorWig_y} to treat each of the $N$ Dirac sectors one at a time. This yields
\begin{equation}
    \frac{2^p\,\mathcal{J}^2}{\lambda}\Bigg[\frac{1}{2}\Tr\bigg(\sigma_a\,e^{\frac{\mu\tau}{2p}\sigma_z}\,\sigma_b\,e^{\frac{\mu(\beta-\tau)}{2p}\sigma_z}\bigg)\Bigg]^p \Bigg[\Tr\bigg(e^{\frac{\mu\beta}{2p}\sigma_z}\bigg)\Bigg]^{N-p}\label{eq:wohp_trace_traverse_ab_sigmas}
\end{equation}
where ``$\sigma_a$'' is $\sigma_x$ if $a=0$ and $\sigma_y$ if $a=1$. As we should expect, the answer depends only on whether $a$ and $b$ are equal, giving us
\begin{equation}\label{eq:wohp_traverse_00}
    Z_0\frac{\mathcal{J}^2}{\lambda}\Bigg[\frac{\cosh\Big(\frac{\mu}{p}\big(\beta/2-\tau\big)\Big)}{\cosh\Big(\frac{\mu\beta}{2p}\Big)}\Bigg]^p
\end{equation}
if $a=b$ and 
\begin{equation}\label{eq:wohp_traverse_01}
    Z_0\frac{\mathcal{J}^2}{\lambda}\Bigg[\frac{\sinh\Big(\frac{\mu}{p}\big(\beta/2-\tau\big)\Big)}{\cosh\Big(\frac{\mu\beta}{2p}\Big)}\Bigg]^p
\end{equation}
if $a\neq b$, where we have defined
\begin{equation}
    Z_0=\Tr\bigg(e^{-\beta\,\mu H_\text{free}}\bigg)=\Bigg[2\cosh\Big(\frac{\mu\beta}{2p}\Big)\Bigg]^N.
\end{equation}
This is exactly in line with the $G\Sigma$ answer from \cite{Maldacena:2018lmt,Sontag:2026iiu}. We get our first new result from considering two contracted pairs. We will only show one example calculation (a $00$-chord crossing a $01$-chord) as the extrapolation is straightforward. The relevant trace is
\begin{align}
    &\Tr\bigg(\marknode{C}{H}_0\,e^{-\tau_1\,\mu H_\text{free}}\,\marknode{D}{H}_0\,e^{-\tau_2\,\mu H_\text{free}}\,\marknode{E}{H}_1\,e^{-\tau_3\,\mu H_\text{free}}\,\marknode{F}{H}_0\,e^{-(\beta-\tau_1-\tau_2-\tau_3)\,\mu H_\text{free}}\bigg)\contractaa[black, thick]{C}{F}{2.5ex}{}\contractaa[black, thick]{D}{E}{1.7ex}{}\\
    =\sum_{I,I'} J_I^2 J_{I'}^2 &\Tr\bigg(\Psi_I^{(0)}\,e^{-\tau_1\,\mu H_\text{free}}\,\Psi_{I'}^{(0)}\,e^{-\tau_2\,\mu H_\text{free}}\,\Psi_{I'}^{(1)}\,e^{-\tau_3\,\mu H_\text{free}}\,\Psi_I^{(0)}\,e^{-(\beta-\tau_1-\tau_2-\tau_3)\,\mu H_\text{free}}\bigg)
\end{align}
where $I,I'$ are multi-indices of length $p$ and $\Psi_I=\psi_{i_1}...\psi_{i_p}$. We consider all $\tau_i\geq 0$ with $\tau_1+\tau_2+\tau_3\leq\beta$. We can again simplify this as a product over the $N$ Dirac sectors
\begin{equation}
    =\sum_{I,I'} J_I^2 J_{I'}^2 \prod_{i=1}^N \begin{cases}
        2\cosh\Big(\frac{\mu\beta}{2p}\Big):&i\notin I,i\notin I'\\
        \cosh\Big(\frac{\mu}{p}\big(\beta/2-(\tau_1+\tau_2+\tau_3)\big)\Big):&i\in I,i\notin I'\\
        i\sinh\Big(\frac{\mu}{p}\big(\beta/2-\tau_2\big)\Big):&i\notin I,i\in I'\\
        -\frac{i}{2}\sinh\Big(\frac{\mu}{p}\big(\beta/2-(\tau_1+\tau_3)\big)\Big):&i\in I,i\in I'
    \end{cases}
\end{equation}
\begin{samepage}
\begin{align}
    =\;&Z_0\bigg(\frac{\mathcal{J}^2}{\lambda}\bigg)^2\Bigg[\frac{\cosh\Big(\frac{\mu}{p}\big(\beta/2-(\tau_1+\tau_2+\tau_3)\big)\Big)}{\cosh\Big(\frac{\mu\beta}{2p}\Big)}\Bigg]^p\,\Bigg[\frac{\sinh\Big(\frac{\mu}{p}\big(\beta/2-\tau_2\big)\Big)}{\cosh\Big(\frac{\mu\beta}{2p}\Big)}\Bigg]^p\nonumber\\&\cdot\mathbb{E}\Bigg[\bigg(-\frac{\sinh\Big(\frac{\mu}{p}\big(\beta/2-(\tau_1+\tau_3)\big)\Big)\cosh\Big(\frac{\mu\beta}{2p}\Big)}{\cosh\Big(\frac{\mu}{p}\big(\beta/2-(\tau_1+\tau_2+\tau_3)\big)\Big)\sinh\Big(\frac{\mu}{p}\big(\beta/2-\tau_2\big)\Big)}\bigg)^{|I\cap I'|}\Bigg]\label{eq:wohp_crossed0010}
\end{align}
\end{samepage}
where the expectation value is taken uniformly over all possible $I,I'$. Equivalently, it can be taken over the hypergeometrically distributed random variable $|I\cap I'|$. This is the exact answer at finite $p,N$ and does not diverge anywhere. We can recover the doubled-scaled answer to leading order in $\lambda$ (given in Appendix A of \cite{Sontag:2026iiu}) by approximating the hypergeometric distribution as Poisson, which is valid as long as the base of $|I\cap I'|$ is $\ll\sqrt{p}$ (i.e. $\tau_2$ is not too close to $\beta/2$).\footnote{There is an intermediate option of using the binomial approximation to the hypergeometric distribution which extends the region of validity up to $\ll p$, but we have found no use for this thus far.} This gives us
\begin{align}
    \approx\;&Z_0\bigg(\frac{\mathcal{J}^2}{\lambda}\bigg)^2\Bigg[\frac{\cosh\Big(\frac{\mu}{p}\big(\beta/2-(\tau_1+\tau_2+\tau_3)\big)\Big)}{\cosh\Big(\frac{\mu\beta}{2p}\Big)}\Bigg]^p\,\Bigg[\frac{\sinh\Big(\frac{\mu}{p}\big(\beta/2-\tau_2\big)\Big)}{\cosh\Big(\frac{\mu\beta}{2p}\Big)}\Bigg]^p\nonumber\\&\cdot\exp\Bigg(-\frac{\lambda}{2}\Bigg[1+\frac{\sinh\Big(\frac{\mu}{p}\big(\beta/2-(\tau_1+\tau_3)\big)\Big)\cosh\Big(\frac{\mu\beta}{2p}\Big)}{\cosh\Big(\frac{\mu}{p}\big(\beta/2-(\tau_1+\tau_2+\tau_3)\big)\Big)\sinh\Big(\frac{\mu}{p}\big(\beta/2-\tau_2\big)\Big)}\Bigg]\Bigg)\label{eq:wohp_crossed0010_Poisson}
\end{align}
which, if we take $\mu\beta>p\ln p$ to enter the cylinder phase (and again keep $\tau_2$ far from $\beta/2$), simplifies considerably to
\begin{equation}
    Z_0\bigg(\frac{\mathcal{J}^2}{\lambda}\bigg)^2\exp\bigg(-\mu\,\min(\tau_{123}\,,\,\beta-\tau_{123})-\mu\,\min(\tau_2\,,\,\beta-\tau_2)-\lambda\,\theta(\beta/2-\tau_1-\tau_3)\theta(\beta/2-\tau_2)\bigg)
\end{equation}
where $\tau_{123}=\tau_1+\tau_2+\tau_3$. This aligns with the simple geometric answer discussed in \cite{Sontag:2026iiu}, where we are told to draw the shortest chord connecting each contracted pair of nodes on the cylinder and then apply a weight of $e^{-\mu\tau}$ for each $\tau$-separation and a weight of $e^{-\lambda}$ for each crossing, independent of chord type. Our matrix representation of the Majoranas gives us an exact answer for the two-point and four-point thermal DSSYK-MQ correlators which in the triple scaling limit (double scaling plus small $\lambda$) reproduces the $G\Sigma$ results.

\section{Adding the heavy particle}\label{sect:whp}

To study the AS$^2$ saddle point, we analyze the heavy two-point function
\begin{equation}\label{eq:whp_Z_AS2}
    Z_{AS^2}[\beta]=\Tr\Big(e^{-\frac{\beta}{2} H_{MQ}}\,\mathbb{O}^{(0)}\;e^{-\frac{\beta}{2} H_{MQ}}\,\mathbb{O}^{(0)}\Big)
\end{equation}
where $\mathbb{O}^{(0)}$ is defined on SYK site $0$ as in \eqref{eq:intro_SYK_matter} with $\Delta \gtrsim \mathcal{O}(1/\lambda)$. For simplicity, we will take $\Delta$ to be an integer. The maximum possible mass is $\Delta_\text{max}=N/p=2p/\lambda$ at which point $\mathbb{O}$ includes all $N$ Majorana fermions.\footnote{Although the matter chord which delivers the largest crossing penalty is half of this mass, $\Delta=\Delta_\text{max}/2=p/\lambda$. This will make some features of the phase diagram in Figure~\ref{fig:sowcn_phase_diagram} less surprising.} We can again expand \eqref{eq:whp_Z_AS2} as a Dyson series
\begin{equation}\label{eq:whp_Dyson_series}
    =\sum_{n=0}^\infty \frac{1}{n!}\sum_{a_i\in\{0,1\}}\int_0^\beta d\tau_1...d\tau_n \Tr\Bigg(\mathcal{T}\bigg(e^{-\beta\,\mu H_\text{free}}\,\mathbb{O}^{(0)}(\beta/4)\,\mathbb{O}^{(0)}(3\beta/4)\,H_{a_1}(\tau_1)...H_{a_n}(\tau_n)\bigg)\Bigg)
\end{equation}
situating the matter operators at Euclidean times $\beta/4$ and $3\beta/4$. We again integrate out the random $J_{i_1...i_p}$ and $K_{i_1...i_{p'}}$ couplings to consider only fully-contracted diagrams. If we take $\mu\beta>\mathcal{O}(p\ln p)$ to go deep into the cylinder phase, we can guarantee that no contracted pair of $H_{a}$'s will span anywhere near $\mathcal{O}(\beta/2)$ Euclidean time without using the $\mathbb{O}$'s. This lets us distinguish between the ``wormhole chords'' (the contractions spanning $\mathcal{O}(\beta/2)$ Euclidean time) and the ``regular chords'' (the rest). We will then subdivide our sum in \eqref{eq:whp_Dyson_series} into sectors labeled by $m$, the number of wormhole chords. The relevant combinatorial factor is then
\begin{equation}\label{eq:whp_mfactorial_squared}
    \frac{1}{n!}\begin{pmatrix}n\\2m\end{pmatrix}\begin{pmatrix}2m\\m\end{pmatrix}=\frac{1}{(n-2m)!}\,\frac{1}{(m!)^2}=\frac{1}{(n_\text{reg})!}\,\frac{1}{(m!)^2}
\end{equation}
where $n_\text{reg}$ is the number of ``regular chord'' nodes.

The regular chords away from the wormhole mouths will follow the same rules as in the standard cylinder phase. We are interested in the behavior of the wormhole chords. To begin analyzing these, we first consider the simplest case where $n_\text{reg}=0$ and all of the $m$ wormhole chords are anchored entirely on boundary $0$, have no Euclidean time between them, and do not cross each other. We define
\begin{equation}\label{eq:whp_def_Zm}
    Z_m[\beta]=\Tr\bigg(e^{-\frac{\beta}{2}\,\mu H_\text{free}}\,\mathbb{O}^{(0)}\,\marknode{G}{H}_0...\marknode{H}{H}_0\,e^{-\frac{\beta}{2}\,\mu H_\text{free}}\,\marknode{I}{H}_0...\marknode{J}{H}_0\,\mathbb{O}^{(0)}\bigg)\contractaa[black, thick]{G}{J}{2.5ex}{}\contractaa[black, thick]{H}{I}{1.7ex}{}
\end{equation}
where each ellipsis indicates $m$ insertions of $H_0$. We can simplify this as before by writing the sum over multi-indices $I_1...I_m$ of length $p$ and multi-index $I'$ of length $p'=p\Delta$. We will also use the fact that $p\equiv 0\text{ mod }4$ to conjugate all of the $\Psi$'s in the right clump to get
\begin{equation}    Z_m[\beta]=\sum_{I',I_1...I_m}K^2_{I'}J^2_{I_1}...J^2_{I_m}\Tr\bigg(e^{-\frac{\beta}{2}\,\mu H_\text{free}}\,\Psi_{I'}^{(0)}\,\Psi_{I_1}^{(0)}...\Psi_{I_m}^{(0)}\,e^{-\frac{\beta}{2}\,\mu H_\text{free}}\,\Psi_{I_m}^{(0)\dagger}...\Psi_{I_1}^{(0)\dagger}\,\Psi_{I'}^{(0)\dagger}\bigg).
\end{equation}
Notice that now the order of Majoranas $\psi_i$ in the left clump is precisely the opposite of that in the right clump. This frees us to commute any Majorana past any other in each clump, as long as we do the mirror thing in the opposite clump so that any signs cancel. We can use this to annihilate matching pairs of Majoranas until each clump can be described as $\Psi_{I''}^{(0)(\dagger)}$ for some multi-index $I''$, with no repeats, whose length depends on all of the $I',I_1...I_m$. This then leaves us with the same computation as in \eqref{eq:wohp_trace_traverse_ab_psis}-\eqref{eq:wohp_traverse_00}, which we know yields
\begin{align}    Z_m[\beta]&=\frac{Z_0}{2^{(m+\Delta)p}}\sum_{I',I_1...I_m}K^2_{I'}J^2_{I_1}...J^2_{I_m}\bigg(\cosh\Big(\frac{\mu\beta}{2p}\Big)\bigg)^{-|I''|}\\
&=Z_0\bigg(\frac{\mathcal{K}^2}{\lambda}\bigg)\bigg(\frac{\mathcal{J}^2}{\lambda}\bigg)^m \;\mathbb{E}\Bigg[\bigg(\cosh\Big(\frac{\mu\beta}{2p}\Big)\bigg)^{-|I''|}\Bigg]\label{eq:whp_Zm_expectation_value}
\end{align}
where now the expectation value is taken uniformly over all possible $I',I_1...I_m$. The quantity $|I''|$ is not distributed hypergeometrically, but instead as a random walk on the space $(\mathbb{Z}_2)^N$ with steps of size $p$. We can Fourier transform this random walk to derive the exact distribution. The details can be found in Appendix~\ref{app:FourierWalk}, with the final answer given in \eqref{eq:appA_final}. Plugging it into the above, we get
\begin{align}
    Z_m[\beta]&=Z_0\bigg(\frac{\mathcal{K}^2}{\lambda}\bigg)\bigg(\frac{\mathcal{J}^2}{\lambda}\bigg)^m\sum_{c=0}^N \bigg(\cosh\Big(\frac{\mu\beta}{2p}\Big)\bigg)^{-c}\nonumber\\
    &\quad\cdot\frac{1}{2^N}\sum_{k=0}^N K_c(k;N)\;K_k(p\Delta;N)\;_2F_1(-p,-k;-N;2)^m\label{eq:whp_Zm_with_c}
\end{align}
within which we recognize the Kravchuck generating series noted in \eqref{eq:appA_Kravchuck_generating}, giving
\begin{align}
    \sum_{c=0}^N K_c(k;N)\bigg(\cosh\Big(\frac{\mu\beta}{2p}\Big)\bigg)^{-c}&=\bigg(1-\sech\Big(\frac{\mu\beta}{2p}\Big)\bigg)^k\bigg(1+\sech\Big(\frac{\mu\beta}{2p}\Big)\bigg)^{N-k}\\
    &= \bigg(\tanh\Big(\frac{\mu\beta}{4p}\Big)\bigg)^{2k}\;\bigg(1+\sech\Big(\frac{\mu\beta}{2p}\Big)\bigg)^N.
\end{align}
Combining all of our constant factors gives us
\begin{equation}\label{eq:whp_Z0tilde}
    \tilde{Z}_0=Z_0\bigg(\frac{\mathcal{K}^2}{\lambda}\bigg)\frac{1}{2^N}\bigg(1+\sech\Big(\frac{\mu\beta}{2p}\Big)\bigg)^N=\bigg(\frac{\mathcal{K}^2}{\lambda}\bigg)\Bigg[\cosh\Big(\frac{\mu\beta}{2p}\Big)+1\Bigg]^N
\end{equation}
so that we are left with
\begin{equation}\label{eq:whp_finalZm}
    Z_m[\beta]=\tilde{Z}_0\bigg(\frac{\mathcal{J}^2}{\lambda}\bigg)^m\sum_{k=0}^N K_k(p\Delta;N)\;\bigg(\tanh\Big(\frac{\mu\beta}{4p}\Big)\bigg)^{2k}{}_2F_1(-p,-k;-N;2)^m.
\end{equation}

There are already a number of things to note about this formula. If $\mu\beta=\mathcal{O}(1)$, then $\tilde{Z}_0 \sim 2^N$, and the $\tanh$ in \eqref{eq:whp_finalZm} isolates the $k=0$ term of the sum, which is $1$. So the action $\ln Z$ is of order $N\ln 2$ independent of $m$ as we should expect in the disks phase.\footnote{This remains true even when we cannot easily distinguish between ``wormhole'' and ``regular'' chords. The heavy $\mathbb{O}$'s simply contract normally across a (pinched) disk.} On the other hand, if $\mu\beta>\mathcal{O}(p\ln p)$, then the $\beta$-independent $1/2$ in $\cosh$ makes $\tilde{Z}_0\sim 2^{-N}$. For sufficiently large $m$, the action is then of order $-N\ln 2$ as we should expect in the AS$^2$ phase! We will see later that for smaller $m$, the sum over $k$ gives another $2^N$ to bring the action back to $\sim0\ln 2$ for the cylinder phase. We can see this here at $m=0$ where another Kravchuck generating series reduces our sum over $k$ to
\begin{align}
    Z_{m=0}[\beta]&=\tilde{Z}_0\bigg(1-\tanh^2\Big(\frac{\mu\beta}{4p}\Big)\bigg)^{p\Delta}\bigg(1+\tanh^2\Big(\frac{\mu\beta}{4p}\Big)\bigg)^{N-p\Delta}\\
    &=\tilde{Z}_0\;2^N\;\bigg(\cosh\Big(\frac{\mu\beta}{2p}\Big)\bigg)^{-p\Delta}.
\end{align}
So already our analysis has born an important insight: the $\chi=-2$ contribution to the action for the AS$^2$ saddle is indeed present in the DSSYK-MQ model, but only once we turn on $\mathcal{J}$ so that we can consider $m>\mathcal{O}(1)$.

\subsection{Summing over wormhole chord number}\label{subsect:summing_over_m}

To get a better handle on the actual form of the wormhole, we need to consider more complicated contraction patterns than \eqref{eq:whp_def_Zm}. For simplicity, we will now take $\mu\beta=\alpha\, p\ln p$ for sufficiently large $\alpha$ such that we can drop all $\beta$ dependence outside of $\tilde{Z}_0$ (our simulations suggest this is around $\alpha\gtrsim 3$). We will keep $n_\text{reg}=0$ but restore the integral over $\tau$ for our wormhole chords. Our previous computations \eqref{eq:wohp_traverse_00}-\eqref{eq:wohp_traverse_01} tell us that each chord node insertion should then come with a factor of\footnote{This is an approximation that ignores the possible drift and dispersion of the wormhole mouth as $m$ exceeds $\Delta$ and wormhole chords start contracting with each other instead of with $\mathbb{O}$. Including this would be a small correction that increases the numerical factor in the numerator, as the rest of the fraction is fixed by units.}
\begin{equation}\label{eq:sowcn_integrate_dtau}
    \frac{\mathcal{J}}{\sqrt{\lambda}}\sum_{a\in\{0,1\}}\int_0^\beta d\tau\,\exp(-\mu|\tau-\beta/4|)\approx\frac{4\mathcal{J}}{\mu\sqrt{\lambda}}.
\end{equation}
On top of this, we need to deal with wormhole chords crossing each other. An exact answer for this including $\beta$ dependence a la \eqref{eq:wohp_crossed0010} would be interesting, but is unnecessary for understanding the bigger picture. We will use the following combinatorial argument to approximate. By taking $\beta$ large, we have forced every Majorana in each $\mathbb{O}$ to be annihilated by nearby wormhole chord insertions. In other words, we are restricting to the $c=0$ sector of \eqref{eq:whp_Zm_with_c}. This means that of the $pm$ total Majoranas in the chord node insertions nearby an $\mathbb{O}$, $p\Delta$ must be used for annihilating that $\mathbb{O}$. If we assume that these $p\Delta$ are on average spread relatively evenly among the $m$ nodes,\footnote{This is only a good assumption when all of the wormhole chords are $00$-type. We comment further on this in section~\ref{sect:discuss}.} then each node has a set of $\approx p\Delta/m$ Majoranas that is guaranteed to have no overlap with any other node's analogous set. Taking the double scaling limit while holding $m/\Delta$ fixed, we find that the total overlap is distributed as
\begin{equation}
    |I_i\cap I_j|\sim \text{Poisson}\Bigg(\frac{p^2}{N}\bigg(1-\Big(\frac{\Delta}{m}\Big)^2\bigg)\Bigg)
\end{equation}
such that the crossing penalty is $\exp(-\lambda_\text{eff})$ for
\begin{equation}
    \lambda_\text{eff}=\lambda\,\bigg(1-\Big(\frac{\Delta}{m}\Big)^2\bigg).
\end{equation}
The weighted sum over all ways of contracting the $m$ wormhole chords then gives us a factor of\footnote{This too is an approximation at infinite $\beta$. Observe that if $c=0$, fully inverting the contraction order is actually a symmetry, so the ``maximally crossed'' case comes without any penalty. We expect this to yield an $O(1)$ symmetry factor here, likely just $2$, which will not affect our results.\label{fn:sowcn_sym}}
\begin{equation}
    [m]_{q_\text{eff}}\,! = \prod_{k=1}^m \frac{1-e^{-k\lambda_\text{eff}}}{1-e^{-\lambda_\text{eff}}}.
\end{equation}
With this, we can now write the full $n_\text{reg}=0$ wormhole partition function
\begin{equation}\label{eq:sowcn_Zwh}
    Z_\text{wh}=\tilde{Z}_0 \sum_{m=0}^\infty \frac{[m]_{q_\text{eff}}\,!}{(m!)^2} \bigg(\frac{16\mathcal{J}^2}{\lambda\mu^2}\bigg)^m \sum_{k=0}^NK_k(p\Delta;N)\;_2F_1(-p,-k;-N;2)^m
\end{equation}
and perform a saddle point analysis. We have checked the following analysis by comparing it to a matrix simulation of \eqref{eq:whp_Zm_expectation_value}. The details are given in Appendix \ref{app:B}.

The first step is to approximate and then Fourier expand the hypergeometric function. In the double scaling limit, for $k$ far from $N/2$ but potentially $\gg p$ we have
\begin{samepage}
\begin{align}
    _2F_1(-p,-k;-N;2)^m&\approx\bigg|1-\frac{2k}{N}\bigg|^{\,pm}\\
    &\approx e^{-pm\Big(\frac{2k}{N}+\frac{2k^2}{N^2}\Big)}+e^{-pm\Big(\frac{2(N-k)}{N}+\frac{2(N-k)^2}{N^2}\Big)}\\
    &=\frac{N}{\sqrt{8\pi p m}}\int_{-\infty}^\infty dy\,e^{-\frac{N^2y^2}{8pm}}\Bigg[\bigg(e^{-\frac{2pm}{N}+iy}\bigg)^k+\bigg(e^{-\frac{2pm}{N}+iy}\bigg)^{N-k}\Bigg].\label{eq:sowcn_Fourier}
\end{align}
\end{samepage}
We also know that the Kravchuck polynomials satisfy $K_k(p\Delta;N)=K_{N-k}(p\Delta;N)$ since $p\Delta$ is even, so we can use a generating series one more time to write
\begin{align}
    Z_\text{wh}\,\propto\,&\sum_{m=0}^\infty\int_{-\infty}^\infty dy \;\Bigg(\frac{[m]_{q_\text{eff}}\,!}{(m!)^2\sqrt{m}} \bigg(\frac{16\,\mathcal{J}^2}{\lambda\,\mu^2}\bigg)^m e^{-\frac{N^2y^2}{8pm}}\bigg[1-e^{-\frac{2pm}{N}+iy}\bigg]^{p\Delta}\bigg[1+e^{-\frac{2pm}{N}+iy}\bigg]^{N-p\Delta}\Bigg)\label{eq:sowcn_full_action}\\
    \equiv\,&\sum_{m=0}^\infty\int_{-\infty}^\infty dy\;\,\exp\Big(\Phi(y,m)\Big)\label{eq:sowcn_def_Phi}
\end{align}
for which we can extremize the action $\Phi(y,m)$. We get a good approximation using $y_*\ll 1$, which yields
\begin{equation}
    y_*\approx \,i\;\frac{z_0(1-2p\Delta/N-z_0)(1-z_0^2)}{\frac{N}{4pm}(1-z_0^2)^2+z_0\big((1-2p\Delta/N)(1+z_0^2)-2z_0\big)}\quad\text{for}\quad z_0=e^{-\frac{2pm}{N}}.\label{eq:sowcn_def_ystar}
\end{equation}

\begin{figure}[t]
\centering
\hspace{-1cm}
\includegraphics[width = 0.8\textwidth]{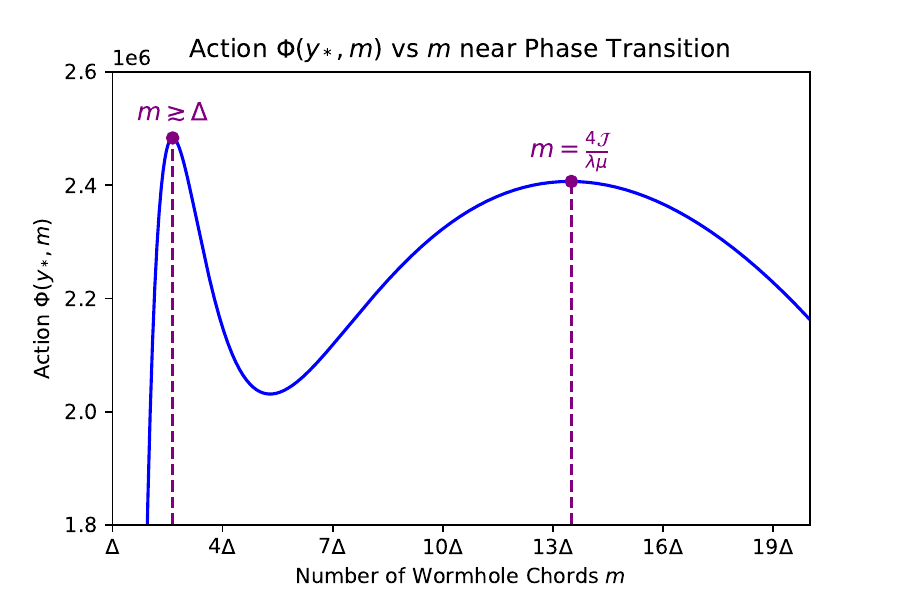}
\caption{Plot of the action $\Phi(y,m)$ from \eqref{eq:sowcn_def_Phi} evaluated at $y=y_*(m)$ from \eqref{eq:sowcn_def_ystar} displaying two local maxima corresponding to the perturbed thermal AdS (left) and AS$^2$ (right) phases. This example is tuned to keep both maxima comparable; in generic snapshots the left peak is either much sharper and closer to $m=\Delta$ or completely drowned out by the right peak. This data uses parameters $N=2\times10^8$, $p=10^3$, $\frac{p\Delta}{N}=\frac{4}{9}$, and $\frac{\mathcal{J}}{\mu}=3\times10^3$ (the $\beta$ dependance is ignored, corresponding to $\mu\beta\gtrsim 3p\ln p$). \label{fig:sowcn_two_extrema}}
\end{figure}

We resort to numerics for finding $m_*$. In the relevant regime of free parameters $p,\lambda,\Delta,\frac{\mathcal{J}}{\mu}$ we get two competing local maxima: one with $m_*$ close to (but greater than) $\Delta$, and the other with $m_*\approx \frac{4\mathcal{J}}{\lambda\mu}$ which is (in this regime) many times larger than $\Delta$. We show an example in Figure~\ref{fig:sowcn_two_extrema}. We interpret the former $m_*\gtrsim\Delta$ as characterizing the perturbed thermal AdS phase without a wormhole, since its sharp peak means that adding additional ``wormhole'' chords comes at a significant entropic penalty, making the holographic bulk distance $\ell$ between the two heavy operator insertions very large.\footnote{It is the same order of magnitude as the bulk ``distance'' between the two disconnected disks in the disks phase.} Also, for the majority of this phase $\lambda_\text{eff}$ is close to zero, so the apparent circumference $\lambda_\text{eff}\,m_*$ of the ``wormhole'' (as measured by a probe crossing through it) is very small.\footnote{Probes not crossing through the ``wormhole'' still see $\lambda$ instead of $\lambda_\text{eff}$. But such a probe inserted at one edge of the mouth and absorbed at the other (a Euclidean time $\tau=\mathcal{O}(1/\mu)$ away) will never see any strange features, since all of the ``wormhole'' Majoranas annihilate internally.} Finally, in this phase the quantity $e^{-\frac{2m_*p}{N}+iy_*}=\mathcal{O}(1)$, so the rightmost factor in \eqref{eq:sowcn_full_action} gives us the extra $2^N$ back that indicates $\chi=0$.


We interpret the latter $m_*\approx \frac{4\mathcal{J}}{\lambda\mu}$ as characterizing the AS$^2$ wormhole phase. The distance through the wormhole is dominated by crossing wormhole chords instead of entropic penalties, and $e^{-\frac{2m_*p}{N}+iy_*}$ is small enough that we retain $\chi=-2$. Since $m_*\gg\Delta$ we have $\lambda_\text{eff}\approx\lambda$, so chord crossing rules inside the wormhole remain largely the same as usual.\footnote{This is at least true for test particles with masses far from $\Delta$ crossing an $\mathcal{O}(1)$ number of wormhole chords. If we try to probe further, we may run into the same issue as discussed in footnote \ref{fn:sowcn_sym}.} This implies that an observer inside the baby universe should experience the same standard semiclassical physics as they would in the disks or cylinder phases. We discuss this further in section \ref{sect:discuss}.

\begin{figure}[t]
\centering
\hspace{-1cm}
\includegraphics[width = 0.7\textwidth]{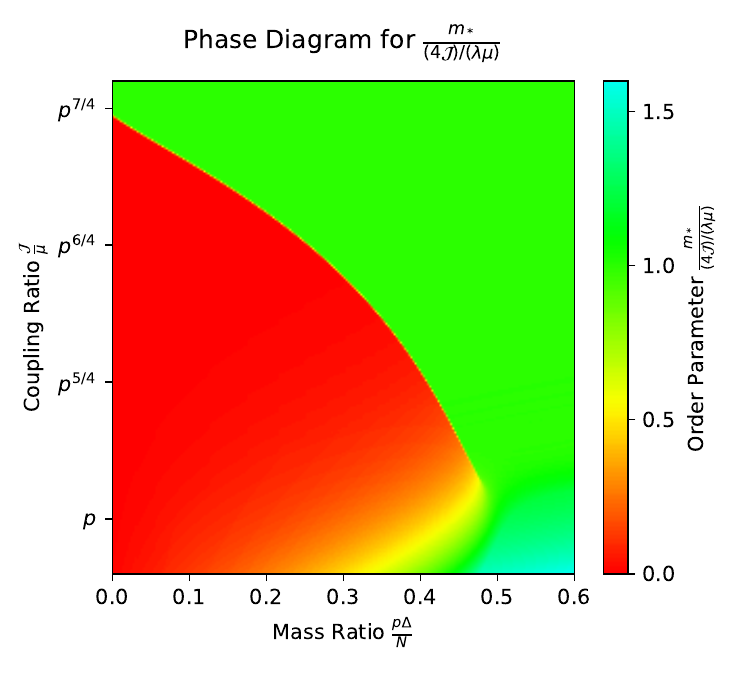}
\caption{Phase diagram showing the order parameter $\frac{m_*}{(4\mathcal{J})/(\lambda\mu)}$ plotted as a function of coupling ratio $\frac{\mathcal{J}}{\mu}$ and mass ratio $\frac{p\Delta}{N}$. The red region shows the perturbed thermal AdS phase (with no wormhole), and the green region shows the AS$^2$ phase (with a wormhole). The phase transition is first order for sufficiently large $\frac{\mathcal{J}}{\mu}$ but has a critical point close to $\frac{\mathcal{J}}{\mu}=p$, $\frac{p\Delta}{N}=\frac{1}{2}$. This data uses parameters $N=2\times10^8$, $p=10^3$ and the same $\mu\beta\gtrsim 3p\ln p$ as in Figure~\ref{fig:sowcn_two_extrema}. \label{fig:sowcn_phase_diagram}}
\end{figure}

The two competing maxima give us a first order phase transition between the perturbed thermal AdS and AS$^2$ phases. To explore this further, we will use the quantity
\begin{equation}
    \frac{m_*}{(4\mathcal{J})/(\lambda\mu)}
\end{equation}
as our order parameter, since it is $\approx 0$ in the perturbed thermal AdS phase and $\approx 1$ in the AS$^2$ phase. A phase diagram is shown in Figure~\ref{fig:sowcn_phase_diagram}.
For sufficiently large coupling ratio $\frac{\mathcal{J}}{\mu}$, we indeed see a sharp first order transition between the two phases. We also see a critical point near half of the maximal mass (i.e. $\frac{p\Delta}{N}=\frac{1}{2}$) at which the transition becomes second order. We find that increasing $p$ and $N$ while keeping $\lambda$ fixed shifts this critical point logarithmically closer to the point on the diagram with coupling ratio $\frac{\mathcal{J}}{\mu}=p$ and mass ratio $\frac{p\Delta}{N}=\frac{1}{2}$.

For coupling ratios below this critical point, there is only one local maximum $m_*$ which grows in magnitude with $\Delta$ as it smoothly interpolates between the perturbed thermal AdS phase and the AS$^2$ phase. It even extrapolates farther, as seen by the small blue patch in the bottom right corner of Figure~\ref{fig:sowcn_phase_diagram} (though this may just be an indication that our choice of order parameter becomes a poor descriptor here). We hypothesize that the bulk geometries along this path look something like thin ``spaghettified'' wormholes, in analogy to the disks/cylinder transition of the unperturbed DSSYK-MQ model (the middle region of the number line in Figure~\ref{fig:intro_beta_number_line}). We leave an analysis of the semiclassical conditions inside these interpolative baby universes to future work.\footnote{It could be the case that bulk physics deviates significantly from semiclassics here, as these saddle points may not be sufficiently ``far from topology-changing conditions.''} Moving forward, we will consider only situations where $\frac{\mathcal{J}}{\mu}$ is sufficiently large to give us a first order transition.

\section{Big bang and big crunch}\label{sect:big_bang_crunch}

To probe the cosmology of this baby universe, we need to evolve a slice of it forward in Lorentzian time. If we think about our partition function $Z_{AS^2}[\beta]$ from \eqref{eq:whp_Z_AS2} as computing the norm of a two-sided quantum state
\begin{equation}
    |\Psi_\mathbb{O}\rangle = e^{-\frac{\beta}{4} H_{MQ}}\,\mathbb{O}^{(0)}\,e^{-\frac{\beta}{4} H_{MQ}}
\end{equation}
then the naive way forward is to define
\begin{equation}
    |\Psi_\mathbb{O}(t)\rangle = e^{-\big(\frac{\beta}{4}-it\big) H_{MQ}}\,\mathbb{O}^{(0)}\,e^{-\big(\frac{\beta}{4}+it\big) H_{MQ}}
\end{equation}
and then to try to invent an operator $\hat{m}$ which counts the number of wormhole chords and compute its expected value
\begin{equation}
    \langle\Psi_\mathbb{O}(t)|\,\hat{m}\,|\Psi_\mathbb{O}(t)\rangle.
\end{equation}
This almost works, and it fails in an informative way. We can find this expected value by deforming our integration contour from \eqref{eq:sowcn_integrate_dtau} so that the heavy operators $\mathbb{O}$ remain at $\tau=\beta/4$ and $\tau=3\beta/4$, but now the cut between bra and ket occurs at $\tau=0-it$ and $\tau=\beta/2+it$. We might anticipate the oscillatory part of this integral to change the magnitude of the effective coupling ratio $\big(\frac{\mathcal{J}}{\mu}\big)_\text{eff}$ and thus the size of the baby universe. But since the function we are integrating is analytic (holding fixed the operator ordering on the contour), this contour deformation does not have any effect on $\big(\frac{\mathcal{J}}{\mu}\big)_\text{eff}$ and thus on $m_*$. This is actually what we should have hoped for, however: Lorentzian time evolution on the boundaries commutes with\footnote{In the $\mu\beta\rightarrow\infty$ limit.} operators like $\hat{m}$ inside the baby universe.

Time in a closed universe must be measured relative to some clock inside that universe, not relative to a disconnected boundary. The only object inside our baby universe is the heavy matter particle $\mathbb{O}$. If we define 
\begin{equation}\label{eq:bbc_Ooft}
    \mathbb{O}(t)\equiv e^{it\,\mu H_\text{free}}\,\mathbb{O}\;e^{-it\,\mu H_\text{free}}
\end{equation}
so that
\begin{equation}
    |\Psi_{\mathbb{O}(t)}\rangle=e^{-\frac{\beta}{4} H_{MQ}}\,e^{it\,\mu H_\text{free}}\,\mathbb{O}\;e^{-it\,\mu H_\text{free}}\,e^{-\frac{\beta}{4} H_{MQ}}
\end{equation}
then we find that the analogous contour is now two disconnected pieces. It runs directly from $\tau=\beta/4+it$ to $3\beta/4+it$ on the right side, and from $\tau=3\beta/4-it$ to $\beta/4-it$ on the left side. So now we give each wormhole chord node insertion a factor of
\begin{equation}\label{eq:bbc_integrate_dtau}
    \frac{\mathcal{J}}{\sqrt{\lambda}}\sum_{a\in\{0,1\}}\Bigg[\int_{0-it}^{\beta/4-it} d\tau\,e^{-\mu\tau}+\int_{0+it}^{\beta/4+it} d\tau\,e^{-\mu\tau}\Bigg]\approx\frac{4\mathcal{J}}{\mu\sqrt{\lambda}}\,\cos(\mu t).
\end{equation}
Each wormhole chord has two nodes and so gets a factor of this quantity squared, which means there is no issue with the cosine becoming negative. We can bake this time dependence into our definition of the effective coupling ratio
\begin{equation}
    \bigg(\frac{\mathcal{J}}{\mu}\bigg)_\text{eff}=\frac{\mathcal{J}}{\mu}\,\Big|\cos(\mu t)\Big|
\end{equation}
which then makes our AS$^2$ saddle point
\begin{equation}\label{eq:bbc_mstar_of_t}
    m_*^{AS^2}(t)=\frac{4\mathcal{J}}{\lambda\mu}\,\Big|\cos(\mu t)\Big|.
\end{equation}

Since the circumference of the baby universe is simply $\lambda_\text{eff}\, m_*\approx\lambda m_*$, this is precisely the functional form that we should expect for a big bang / big crunch patch of AdS$_2$ (analogous to equation (1.1) in \cite{Antonini:2023hdh}).\footnote{We can put the metric into the exact form used in \cite{Antonini:2023hdh} if we allow ourselves to define the proper time experienced by the heavy particle $\mathbb{O}$ to be proportional to $t$ with a well-chosen proportionality constant.} And on top of that, our phase diagram in Figure~\ref{fig:sowcn_phase_diagram} tells us how the big bang and big crunch are resolved in the UV complete theory: they are first order phase transitions into and out of the baby universe phase, respectively. The exact lifespan of such a baby universe depends on both $\frac{\mathcal{J}}{\mu}$ and $\Delta$, but it is generically very close to the full $t_\text{max}=\frac{\pi}{\mu}$. Since the vertical axis of Figure~\ref{fig:sowcn_phase_diagram} is log-scaled in powers of $p$, the values of $t$ at which the transitions occur go to $\pm\frac{\pi}{2\mu}$ as $p\rightarrow\infty$.\footnote{Assuming the $p$-scaling of $\frac{\mathcal{J}}{\mu}$ is maintained in the limit.} And the equation \eqref{eq:bbc_mstar_of_t} is not constrained to a single period in $t$, so in some sense this baby universe could also be considered a cyclic ``bouncing'' cosmology---if we can stretch the definition of ``bounce'' to include the process in which a universe ceases to exist and then promptly reemerges.

\section{Discussion}\label{sect:discuss}

In this paper we analyzed the emergence and properties of the AS$^2$ baby universe realized in \mbox{DSSYK-MQ}. We made combinatorial arguments about Majorana traces to get results that remain valid to all orders in $\mathcal{J}$ and $\lambda$. With this we found a first order phase transition between saddle points with and without a baby universe as we varied the mass of the heavy operator $\Delta$ and the coupling ratio $\frac{\mathcal{J}}{\mu}$. We then distinguished between time evolution on the asymptotic thermal AdS boundaries and (a proposal for) time evolution relative to the heavy particle $\mathbb{O}$ inside the baby universe. We showed that the former commutes with the chord structure of the baby universe, while the latter generates a cosmological scale factor proportional to $|\cos(\mu t)|$. This is the exact scale factor that we should expect in a big bang / big crunch patch of AdS$_2$. These findings provoke many interesting follow-up questions, of which we will now dive into two.

\subsection{Conditions inside the baby universe}
First, what can we say about the semiclassical conditions inside the baby universe? The naive way to probe this is to insert some particles with masses $\delta_i=\mathcal{O}(1)$ on one side of the wormhole and then to absorb them on the other side. Our arguments in section \ref{subsect:summing_over_m} tell us that these particles should experience a very normal local geometry with very normal crossing rules. Every crossing between chords of masses $\delta_i$ and $\delta_j$ comes with the usual weight $e^{-\lambda_\text{eff}\,\delta_i\delta_j}$, where Hamiltonian chords have $\delta=1$. This $\lambda_\text{eff}$ depends weakly on both time and the total amount of matter inserted as
\begin{equation}
    \lambda_\text{eff}(t,\delta_\text{tot})=\lambda\Bigg(1-\bigg(\frac{\Delta}{\frac{4\mathcal{J}}{\lambda\mu}|\cos(\mu t)\,|+\delta_\text{tot}}\bigg)^2\Bigg)
\end{equation}
which as discussed above is very close to $\lambda$ for the vast majority of the baby universe's lifetime as $p\rightarrow \infty$. The large number of wormhole chords means that most\footnote{Fitting $m_*\sim \frac{\mathcal{J}}{\lambda\mu}$ wormhole chords into an $\mathcal{O}(\frac{1}{\mu})$ $\tau$-band means a typical distance between wormhole chords of order $\frac{\lambda}{\mathcal{J}}\ll 1$. So any matter particles inserted an $\mathcal{O}(1)$ $\tau$-distance away from $\mathbb{O}$ will have a vanishingly small amplitude of ever seeing it.} matter particles will never run into the heavy operator $\mathbb{O}$ or traverse a meaningful portion of the baby universe's circumference, so they never experience anything out of the ordinary compared to the well-studied disk geometry. And there is no ``inherent length'' or ``throat chord'' to the wormhole in this phase; the distance $\ell$ these particles travel is fully generated by all of the wormhole chords they cross on their way from end to end. Finally, these conditions do not depend sensitively on $N$ or $p$, so observables formed from them should have well-defined and non-erratic double scaling limits and should therefore give rise to good local semiclassics.

The global chord geometry of the baby universe is less immediately clear. Arguments based on a hypothesized dual with JT gravity in \cite{Sasieta:2025vck} as well as the $G\Sigma$ results in \cite{Sontag:2026iiu} strongly suggest that spatial slices of the baby universe are circles. If we declare that the heavy $\mathbb{O}$ worldline is uncrossable, then these circles effectively become intervals, at which point we can use the standard algebraic techniques from \cite{Lin:2023trc} to define the chord Hilbert space. We think it is likely that this suggestion could be confirmed by a careful treatment of crossing rules involving more than just $00$-type wormhole chords. It is true, for instance, that $H^{(1)}_{SYK}$ commutes with both $H^{(0)}_{SYK}$ and $\mathbb{O}^{(0)}$, so each wormhole mouth is in some sense double-sheeted near its center. We are wary, however, of how the large-$\beta$ constraint that every Majorana in the wormhole must appear an even number of times may affect this answer. For example, as mentioned in footnote \ref{fn:sowcn_sym}, taking any baby universe chord diagram and fully inverting the order of nodes at one mouth and not the other will always leave the total crossing penalty unchanged. There is likely more insight to be gained on this point from a closer study of the analogous properties of the unperturbed cylinder geometry (or of the setup analyzed in \cite{bettaque2026sykwormholemicroscope}, but in a heavier $\Delta$ and/or larger $\beta,
\beta'$ phase), but we leave this to future work.

\subsection{Understanding the random couplings}
Another interesting question is how to understand the role of the heavy matter random couplings $K_{i_1...i_{p'}}$ in the emergence of the baby universe. It seems likely that integrating out these Gaussian variables is a necessary step for our construction as opposed to a simplifying approximation: diagrams where the $\mathbb{O}$'s are not contracted to each other should be naturally suppressed by a factor of $\begin{psmallmatrix}N\\p\Delta\end{psmallmatrix}$, but this is maximally $\begin{psmallmatrix}N\\N/2\end{psmallmatrix}\approx\frac{2^N}{\sqrt{N}}$ which is less than the $2^N$ topological suppression for the wormhole. One way to interpret a manual integrating-out is as an average over an ensemble of boundary theories, one for each draw of the $K_{i_1...i_{p'}}$. Ordinarily, such a procedure in SYK involves integrating out $J_{i_1...i_{p}}$ as well. We think the standard self-averaging arguments may make this unnecessary here, but we leave a proof of this to future work. As usual, this interpretation could also be framed as entanglement between the $K_{i_1...i_{p'}}$ and some external reference system which we trace out when performing our calculations \cite{Almheiri:2021jwq}.

Another option is to look at integrating out the $K_{i_1...i_{p'}}$ as a way to generate a large amount of entanglement entropy\footnote{Understood perhaps in the sense of the GNS-constructed microscopic $\alpha$-states discussed in \cite{bettaque2026sykwormholemicroscope}} between the two sides of the thermal circle, which can then be used as a resource for quantum teleportation. The low temperature trace correlates the multi-index $I'$ of each $\mathbb{O}$ with the set of multi-indices $I_1...I_m$ of its nearby $H_{SYK}$ insertions. We then project onto the ``maximally entangled'' state of the $I'$ indices on each side by summing over all cases where they match, which teleports the information in one side's $H_{SYK}$ insertions to those on the other side. 
It is unclear to us, however, what processes would ``deplete'' this teleportation resource, since the number of wormhole chords can increase without bound as $\frac{\mathcal{J}}{\mu}$ increases at fixed $\Delta$.

As a third option, we could simply remove the sum over $I'$ and instead set
\begin{equation}
    \mathbb{O}_\text{fixed}=2^{\frac{p\Delta}{2}}\frac{\mathcal{K}}{\sqrt{\lambda}}\;\psi_1\psi_2...\psi_{p\Delta}.
\end{equation}
Since the rest of the DSSYK-MQ model is fully symmetric under permutations of the Majoranas, this would not change any of the above results. Observers who can only access permutation-symmetric operators would not be able to measure this symmetry breaking. We could then interpret such an $\mathbb{O}_\text{fixed}$ from a third-quantized perspective as the creation operator of a particular baby universe microstate labeled by its list of open Majoranas, here $\{1,2,...,p\Delta\}$. Chords passing through the wormhole scramble this microstate by coherently annihilating or adding Majoranas, and the large-$\beta$ trace projects onto the unique microstate with zero remaining Majoranas. From this perspective, integrating out the $K_{i_1...i_{p'}}$ as above corresponds to putting the baby universe initially into a ``microcanonical'' band which is maximally mixed within a fixed-size sector.

\subsection{Comparison with recent results}
Much of the microscopic analysis performed in \cite{bettaque2026sykwormholemicroscope} is done using the same diagonalized random walk on $(\mathbb{Z}_2)^N$ as we use here. While we consider the low-temperature limit of a wormhole connecting opposite sides of a single thermal circle split between two coupled SYK sites, they instead look at a double trace correlator with a wormhole connecting the two thermal circles of a single SYK site. This allows them to provide a cleaner discussion of closed universe Cauchy slices and their Hilbert space, but obscures the phase transition and Lorentzian cosmology made manifest by the original AS$^2$ setup. By varying $\beta$, $\Delta$, and $\mathcal{J}/\mu$ our model can access all three topologically distinct phases (disks, cylinder, and wormhole) with their correct characteristic entropies. It would be interesting to see whether there exists a limit of our model that aligns with a limit of their model, so that we could apply their Hilbert space results to our AR puzzle \cite{Antonini:2024mci}. One naive guess might be a modified $\beta\rightarrow\infty$ limit of our model, since infinite Euclidean time evolution should be equivalent to a simple ground state projection on both sides of our boundary PETS state, which would cause our un-averaged single $J_{i_1...i_p},K_{i_1...i_{p'}}$ realization to factorize into two pieces which could be associated with their two traces. But then what is the analog of our proposed interior Lorentzian time evolution? It would also be interesting to explore the differences between their model and the cylinder geometry whose chord Hilbert space is discussed in \cite{Sontag:2026iiu}. Can the single heavy particle supporting the wormhole be faithfully replicated by a sufficiently-localized time-dependent $\mu(\tau)$ parameter as in Appendix D of \cite{Sontag:2026iiu}? We hope to return to these questions in future work.

\section*{Acknowledgments}

 We thank Herman Verlinde for helpful discussion. This research is supported by the National Science Foundation Graduate Research Fellowship Program under Grant No. DGE-2444107.

\pagebreak

\appendix
\section{Diagonalizing the random walk on $(\mathbb{Z}_2)^N$}\label{app:FourierWalk}

Given a uniformly random selection of multi-indices $I_1...I_m$ of size $p$ and $I'$ of size $p'=p\Delta$, we are tasked with finding the probability distribution of $|I''|$, the number of indices appearing in $I',I_1...I_m$ an odd number of times. We can compute this using vectors $v$ in the space $(\mathbb{Z}_2)^N$, which contains all binary strings of length $N$. An important measure on this space is the Hamming weight $||v||$, which returns the number of nonzero entries (i.e. $1$'s) in $v$.

We start with a vector $v_0$ with Hamming weight $||v_0||=p\Delta$, representing the multi-index $I'$. We then add vectors $v_1$ through $v_m$ with Hamming weights $||v_i||=p$ to $v_0$ to get a final vector $v_f$. Since this addition is done modulo $2$, the Hamming weight of $v_f$ is precisely the quantity $|I''|$. A uniformly random draw of the $I',I_1...I_m$ corresponds to a uniformly random draw of the $v_0...v_m$ restricted to their respective Hamming weight classes.

The Fourier transform we are interested in takes the form
\begin{equation}
    \tilde{v}(w)=\mathbb{E}\Big[(-1)^{w\cdot v}\Big]
\end{equation}
where the expectation value is taken over the probability distribution in $v$, and the dot product simply counts the number of $1$'s shared by $v$ and $w$. This object factors under addition in $v$ such that
\begin{align}
    \tilde{v}_f(w)&=\mathbb{E}\Big[(-1)^{w\cdot (v_0+v_1+...v_m)}\Big]\\
    &=\mathbb{E}\Big[(-1)^{w\cdot v_0}(-1)^{w\cdot v_1}...(-1)^{w\cdot v_m}\Big]\\
    &=\mathbb{E}\Big[(-1)^{w\cdot v_0}\Big]\mathbb{E}\Big[(-1)^{w\cdot v_1}\Big]...\mathbb{E}\Big[(-1)^{w\cdot v_m}\Big]\\
    &=\mathbb{E}\Big[(-1)^{w\cdot v_0}\Big]\bigg(\mathbb{E}\Big[(-1)^{w\cdot v_1}\Big]\bigg)^m
\end{align}
since the $v_1...v_m$ are all drawn independently but identically. As we should expect, the quantity $\mathbb{E}\Big[(-1)^{w\cdot v_i}\Big]$ depends only on the Hamming weights of the vectors $w$ and $v_i$. Their dot product is hypergeometrically distributed, so we can write\footnote{Going from \eqref{eq:appA_2F1bad} to \eqref{eq:appA_2F1good} requires a specialized Pfaff transformation which can be found at DLMF (\href{https://dlmf.nist.gov/15.8.E7}{15.8.7}).}
\begin{align}
    \mathbb{E}\Big[(-1)^{w\cdot v_i}\Big]&=\sum_{r=0}^N \frac{\begin{pmatrix}||v_i||\\r\end{pmatrix}\begin{pmatrix}N-||v_i||\\||w||-r\end{pmatrix}}{\begin{pmatrix}N\\||w||\end{pmatrix}}(-1)^r\\
    &=\frac{\begin{pmatrix}N-||v_i||\\||w||\end{pmatrix}}{\begin{pmatrix}N\\||w||\end{pmatrix}}\;{}_2F_1\Big(-||v_i||,\,-||w||;\,N-||v_i||-||w||+1;\,-1\Big)\label{eq:appA_2F1bad}\\
    &={}_2F_1\Big(-||v_i||,\,-||w||;\,-N;\,2\Big)\label{eq:appA_2F1good}
\end{align}
and thus have $\tilde{v}_f(w)$ entirely in terms of known functions. 

We can invert this Fourier transform by observing that
\begin{equation}\label{eq:appA_Finv}
    \frac{1}{2^N}\sum_w (-1)^{w\cdot v}=\delta_{v,0}
\end{equation}
so
\begin{align}
    \frac{1}{2^N}\sum_w \tilde{v}_f(w)\,(-1)^{-w\cdot v}
    &=\frac{1}{2^N}\sum_w \mathbb{E}\Big[(-1)^{w\cdot v_f}\Big]\,(-1)^{-w\cdot v}\\
    &=\sum_{v'}P(v_f=v')\;\frac{1}{2^N}\sum_w (-1)^{w\cdot (v'-v)}\\
    &=P(v_f=v).
\end{align}
Since we only care about the Hamming weight $||v_f||$, we can write
\begin{align}
    &P(||v_f||=c)=\sum_{||v||=c}P(v_f=v)\\
    &=\sum_{||v||=c}\;\frac{1}{2^N}\sum_w \tilde{v}_f(w)\;(-1)^{-w\cdot v}\\
    &=\frac{1}{2^N} \sum_{k=0}^N\sum_{\substack{||v||=c\\||w||=k}} (-1)^{w\cdot v}\;\mathbb{E}\Big[(-1)^{w\cdot v_0}\Big]\bigg(\mathbb{E}\Big[(-1)^{w\cdot v_1}\Big]\bigg)^m\\
    &=\frac{1}{2^N}\sum_{k=0}^N\; \begin{pmatrix}N\\c\end{pmatrix}\begin{pmatrix}N\\k\end{pmatrix}{}_2F_1(-k,-c;-N;2)\;{}_2F_1(-p\Delta,-k;-N;2)\;{}_2F_1(-p,-k;-N;2)^m.
\end{align}
The combination of a binomial and hypergeometric function seen twice here is called a Kravchuk polynomial
\begin{equation}\label{eq:appA_def_Kravchuk}
    K_n(x;N)=\begin{pmatrix}N\\n\end{pmatrix}{}_2F_1(-x,-n;-N;2)
\end{equation}
which lets us write the above $m$-dependent probability as
\begin{equation}\label{eq:appA_final}
    P_m(|I''|=c)=\frac{1}{2^N}\sum_{k=0}^N K_c(k;N)\;K_k(p\Delta;N)\;_2F_1(-p,-k;-N;2)^m.
\end{equation}
This is the equation we use to go from \eqref{eq:whp_Zm_expectation_value} to \eqref{eq:whp_Zm_with_c} above. Additionally, the Kravchuck polynomials have the helpful generating series
\begin{equation}\label{eq:appA_Kravchuck_generating}
    \sum_{n=0}^N K_n(x;N)\,z^n=(1-z)^x(1+z)^{N-x}
\end{equation}
which we use multiple times throughout section \ref{sect:whp}.

\section{Markov random walk model}\label{app:B}

We seek here to model the biased random walk analyzed in Appendix~\ref{app:FourierWalk} in order to compute the expectation value piece of $Z_m[\beta]$ from \eqref{eq:whp_Zm_expectation_value}. The difficulty here is in dealing with large values of $\cosh\Big(\frac{\mu\beta}{2p}\Big)$ which cause incredibly rare walks to dominate the naive ensemble. To address this, we consider a Markov process whose unnormalized state vectors encode amplitudes of reaching different values of $|I''|\equiv c$. The transition matrix models adding a new multi-index $I_{i+1}$ to the list $I',I_1...I_i$ and biases each step separately with powers of $\cosh$. This gives us
\begin{equation}
    M(c\rightarrow c')=\frac{\begin{pmatrix}c\\\frac{p+c-c'}{2}\end{pmatrix}\begin{pmatrix}N-c\\\frac{p-c+c'}{2}\end{pmatrix}}{\begin{pmatrix}N\\p\end{pmatrix}}\bigg(\cosh\Big(\frac{\mu\beta}{2p}\Big)\bigg)^{c-c'}\label{eq:appB_transition_matrix}
\end{equation}
which is understood to vanish whenever one of the binomial functions is ill-defined. Our initial state vector $u_0\in (\mathbb{Z}_{\geq 0})^N$ has a $1$ in the position corresponding to $c=p\Delta$ and $0$'s elsewhere. We can then compute the vector $u_m=(M^m)u_0$ and sum all of its components to get the expectation value we want as a function of $m$. If we call this function
\begin{equation}
    F(m)=\mathbb{E}_m\Bigg[\bigg(\cosh\Big(\frac{\mu\beta}{2p}\Big)\bigg)^{-|I''|}\Bigg]
\end{equation}
then it turns out the easier thing to compute and compare is actually
\begin{equation}
    f(m)=\ln\bigg(\frac{F(m+1)}{F(m)}\bigg)
\end{equation}
due to the large exponents involved.

\begin{figure}[h]
\centering
\hspace{-1cm}
\includegraphics[width = \textwidth]{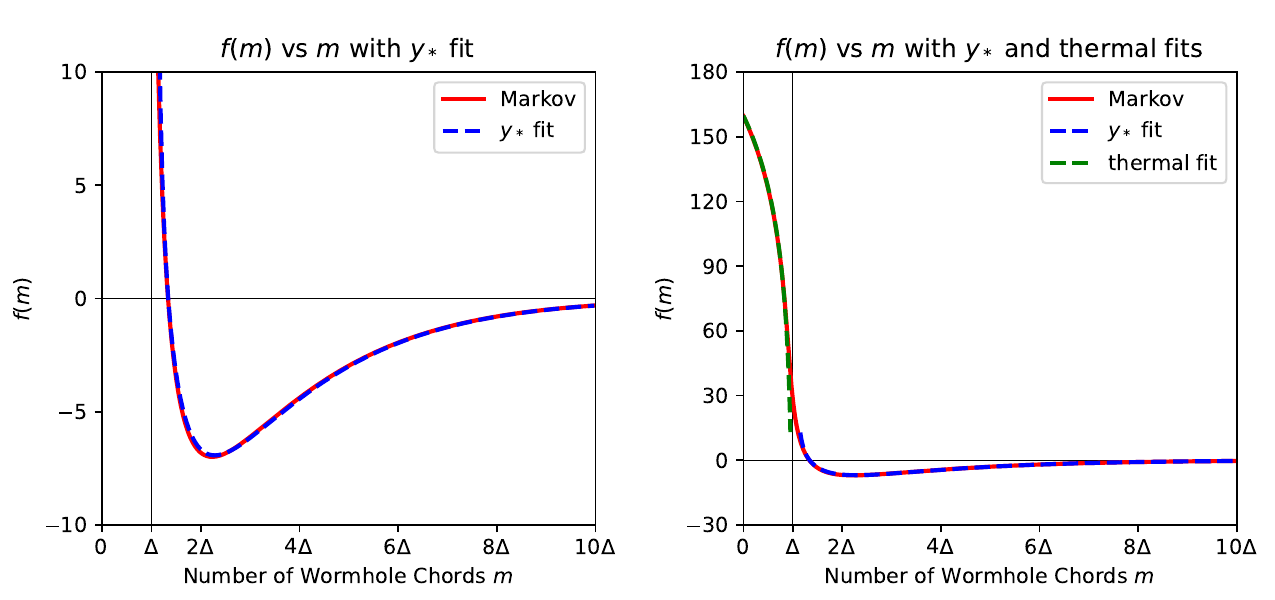}
\caption{Plots comparing the function $f(m)$ computed using the Markov process transition matrix (red lines) to that computed using the $y_*$ saddle point method from \eqref{eq:sowcn_full_action}-\eqref{eq:sowcn_def_ystar} (blue dashed lines) and the thermal approximation \eqref{eq:appB_thermal_fit} (green dashed line). This data uses parameters $N=16384$, $p=48$, $\Delta=85$, and $\mu\beta=2.8\,p\ln p$. The left panel is a zoom-in of the right panel to show the pertinent agreement near $f(m)=0$. \label{fig:appB_Markov_comparison}}
\end{figure}

We find using this method that the saddle point approximation given in \eqref{eq:sowcn_full_action}-\eqref{eq:sowcn_def_ystar}\footnote{Keeping around the factor of $\tanh^2\Big(\frac{\mu\beta}{4p}\Big)$ from \eqref{eq:whp_finalZm}, which is straightforward to pull through the Kravchuck generating series used to get to \eqref{eq:sowcn_full_action}.} is very accurate for $m>\Delta$. Importantly, this includes where $f(m)=0$, which is a left bound on the extrema of $\Phi(y_*,m)$. A comparison is given in the left panel of Figure~\ref{fig:appB_Markov_comparison}. This region is roughly independent of the inverse temperature $\beta$ when $\mu\beta\gtrsim 3p\ln p$. For $m<\Delta$, we find strong temperature dependence which is well approximated by including only the terms in \eqref{eq:appB_transition_matrix} corresponding to $c'=c-p$ and then using Stirling's formula. This yields the ``thermal fit''
\begin{equation}\label{eq:appB_thermal_fit}
    f(m<\Delta)\approx p\,\ln\Bigg(\frac{p(\Delta-m)}{N}\cosh\Big(\frac{\mu\beta}{2p}\Big)\Bigg).
\end{equation}
shown in the right panel of Figure~\ref{fig:appB_Markov_comparison}.


Performing this biased random walk matrix calculation gives us very accurate data for $f(m)$ and therefore $F(m)$. Unfortunately, it is computationally expensive in a way that scales poorly with $p$ and $N$, and we have made little headway distilling from it any kind of unifying approximation. It also does not allow us to take a strict $\mu\beta\rightarrow\infty$ limit, since the transition matrix would diverge. Thankfully, it seems that the $m>\Delta$ regime contains both the perturbed thermal AdS and AS$^2$ saddle points for the parameter ranges that interest us, so we are safe to stick with the $y_*$ approximation.


\printbibliography

\end{document}